\documentclass{aa}  

\usepackage{graphicx}
\usepackage{txfonts}
\newcommand{\hochpunkt}[1]{\mbox{$^{\raisebox{.3ex}{\scriptsize #1}}_{\raisebox{.6ex}{\hspace{.17em}.}}$}}
\newcommand{\hoch}[1]{\mbox{$^{\raisebox{.3ex}{\scriptsize #1}}$}}
\begin{document}

   \title{Time lags of the flickering in cataclysmic variables as a
          function of wavelength}

   \author{Albert Bruch}

   \institute{Laborat\'orio Nacional de Astrof\'{\i}sica,
              Rua Estados Unidos 154, CEP 37504-364, Itajub\'a,
              Brazil\\
              \email{albert@lna.br}
             }

   \date{Received <date> / Accepted <date> }

  \abstract{Flickering is a ubiquitous phenomenon in cataclysmic variables 
(CVs). Although the underlying light 
source is one of the main contributors to the optical radiation, the mechanism 
leading to flickering is not understood as yet.}
{The present study aims to contribute to the set of boundary conditions, 
defined by observations, which must be met by physical models that describe the
flickering. In particular, time lags in the occurrence of flickering events
at different wavelengths over the optical range are examined.}
{To this end, the cross-correlation functions (CCFs) of numerous
light curves of a sample of CVs are analysed that were observed 
simultaneously or quasi-simultaneously in different bands of various 
photometric systems.}
{Deviations of the maxima of the CCFs from zero time-shift indicate a 
dependence of the flickering activity on the wavelength in the sense that
flickering flares reach their maxima slightly earlier in the blue
range than in the red. While the available observational material
does not permit detecting this individually in all observed systems, the 
ensemble of all data clearly shows this effect. Particularly instructive
are the cases of V603~Aql and TT~Ari, where time lags of 15\hochpunkt{s}1 
and 4\hochpunkt{s}3, respectively, are observed between the $U$ and $R$ bands. 
In principle this can be understood if during the development of a flickering
flare the radiation characteristics of the light source responsible for
flickering change 
such that in the early phases of a flare more short-wavelength radiation is 
emitted, and later on, the peak of the emission shifts to the red. Respective
scenarios are discussed and shown to be in qualitative and quantitative
agreement with observations.}{} 

   \keywords{Stars: binaries: close -- Stars: variables: general -- 
Stars: Novae, cataclysmic variables}

\titlerunning{Time shifts of the flickdering in cataclysmic variables}
   \maketitle
%

\section{Introduction}
\label{Introduction}

Cataclysmic variables (CVs) are well known to be short-period interacting 
binary systems where a Roche-lobe filling star, the secondary, transfers 
matter via an accretion disk to a white dwarf primary. One of the most
striking photometric characteristics common to all CVs are short-term
variations of the optical light occurring on time scales of seconds to a 
few dozen minutes with amplitudes of some hundredths of a magnitude up to
more than an entire magnitude. This phenomenon is called flickering and appears
to the eye as a continuous sequence of overlapping flares in the light 
curves of virtually all CVs (but note that this does not necessarily mean 
that flickering consists of a linear superposition of independent events). 
Indeed, it can safely be said that if it does not flicker, is is no CV. 

Flickering has been observed for more then one and a half centuries: In 1856
Pogson
(\cite{Pogson57}) visually observed variations of the dwarf nova U~Gem that
might be interpreted as flickering. So did Baxendell in 1858 (Turner, 
\cite{Turner}). Modern observations of flickering started in the middle
of the twentieth$^{\rm }$ century (Henize \cite{Henize}, Linnell \cite{Linnell}, 
Lenouvel \& Daguillon \cite{Lenouvel}, Johnson et al.\ \cite{Johnson},
Walker \& Herbig \cite{Walker54}, Walker \cite{Walker57}).
The amplitude of the flickering immediately shows that the 
underlying light source is a major contributor to the optical light of the
respective systems. On the mean over all CVs, it emits
at least 25\% of the radiation of
the primary but can easily contribute as much as 50\% and in strong flares
can outshine the non-flickering light sources even by a factor of almost 2
(Bruch \cite{Bruch89}, \cite{Bruch92}). Even though flickering is thus due to 
one of the optically most luminous light sources, it 
remains among the least understood phenomena observed in CVs. 

Flickering is not restricted to CVs or to optical wavelengths. The phenomenon
appears rather to be intimately related to accretion of matter from a disk
onto a central body. Herbst \& Shevchenko (\cite{Herbst}), for instance, 
interpreted flickering-like variations in pre-main-sequence variables as being
due to unsteady accretion (see also Kenyon et al.\ \cite{Kenyon}). A 
comprehensive review of aperiodic X-ray variability in X-ray binaries with
neutron stars and black holes was presented by van der Klis (\cite{vanderKlis}).
Flickering-like variability also occurs in active galactic nuclei [see e.g.\
Garcia et al.\ (\cite{Garcia}) and references therein].
The time scale of the flickering-type variations depends on the dimensions of
the underlying system, and the frequency range where it is best observed 
depends on the compactness of the accretor.

The first encompassing systematic study of the flickering
properties in cataclysmic variables has been performed more than two decades ago by Bruch 
(\cite{Bruch89}, \cite{Bruch92}), who defined several
statistical parameters to describe the phenomenon. In addition
to quantifying 
the already mentioned strong contribution of the flickering light 
source to the total optical light, he showed that the increase of its spectrum 
to shorter wavelength is steeper than can be explained by simple models such
as a black body or a power law. Investigating various physical scenarios, 
Bruch (\cite{Bruch89}, \cite{Bruch92}) concluded that instabilities in the 
inner accretion disk and/or the boundary layer between the disk and the white 
dwarf are most likely to give rise to the observed phenomena.
 
Later work on flickering includes 
attempts to identify the exact location where it 
occurs in CV systems. Already in 1985, Horne \& Stiening (\cite{Horne})
had analysed the variations of RW~Tri through eclipse and concluded that
the flickering light source has the same centre as the accretion disk, but
is less extended. This was later confirmed by Bennie et al.\ (\cite{Bennie}).
Similarly, Welsh et al.\ (\cite{Welsh}) found that flickering in HT~Cas is 
concentrated to the centre of the disk, but is not confined to the immediate
vicinity of the white dwarf. Modifying the technique introduced by Horne \&
Stiening (\cite{Horne}) to avoid systematic effects, Bruch 
(\cite{Bruch96}, \cite{Bruch00}) found similar results for several more 
eclipsing CVs, but that to a lesser degree the hot spot also contributes. 
However, in IP~Peg the hot-spot flickering dominates. In V2051~Oph,
Baptista \& Bortoletto (\cite{Bap04}) identified two flickering components, 
namely a slowly varying one that they associated with the disk overflowing gas 
stream from the secondary star, and a more rapidly varying one with a radial
distribution equal to the steady disk light.

The behaviour of flickering in the frequency domain has been 
the subject of many studies. Fritz \& Bruch (\cite{Fritz}) studied wavelet
transforms of numerous light curves of many CVs. They parametrized the
resulting scalegrams (Scargle et al. \cite{Scargle}) in terms of its
inclination $\alpha$ (frequency behaviour) and its value $\Sigma$ (flickering 
strength) at a reference time scale. For a given system, $\alpha$ and $\Sigma$ 
are stable over many years. On average, 
flickering is somewhat bluer on short time-scales than on longer ones. CVs of 
different types and photometric states occupy distinct regions in the
$\alpha$ -- $\Sigma$ -- plane. A similar study limited to the intermediate
polar V709~Cas was published by Tamburini et al.\ (\cite{Tamburini}).
A wavelet study in X-rays was performed
by Anzolin et al.\ (\cite{Anzolin}). The X-ray data are distributed in a much 
smaller area of the $\alpha - \Sigma$ parameter space than the optical data.
The authors explain the similarity of the X-ray flickering in objects of 
different classes together with the predominance of a persistent stochastic 
behaviour in terms of magnetically driven accretion processes acting in a 
considerable fraction of the analysed objects.

Many studies of the frequency characteristics focus on power spectra where 
flickering light curves in general exhibit a behaviour similar to red noise,
that is, on a double logarithmic scale, a linear decrease of power
towards high frequencies (but with a flattening towards low frequencies;
Bruch \cite{Bruch89}, \cite{Bruch92}; Schimpke \cite{Schimpke}).
Dobrotka et al.\ (\cite{Dobrotka14}) detected two red-noise and two white-noise
components in the X-ray power spectrum of RU~Peg, indicating the presence
of two turbulent regions. The long and low-noise high-cadence light curves 
of some CVs observed by the Kepler satellite represent superb data from which to study
flickering. Scaringi et al.\ (\cite{Scaringi12a}, \cite{Scaringi12b}) 
used these data to construct
the flickering power spectrum of MV~Lyr, which they decomposed into a broken
power law and a series of Lorentzian components, following Belloni et al.\
(\cite{Belloni}) and Novak et al.\ (\cite{Novak}). They detected a linear
rms -- flux relation such that the light source becomes more variable when
it becomes brighter, as has earlier been observed in X-ray binaries and AGNs
(Uttley \& ${\rm M^c}$Hardy \cite{Uttley01}, Uttley et al. \cite{Uttley05},
Heil \& Vaughan \cite{Heil}). This is interpreted as an indication that
flickering variations are due to a multiplicative as opposed to an additive
process, such as shot noise, since in the latter case the rms -- flux relation
would be destroyed. In addition to MV~Lyr, rapid variations were also analysed
in Kepler data of other CVs by Scaringi et al.\ (\cite{Scaringietal14}).

Interpretations of the frequency behaviour in terms of the propagation of
disturbances in the accretion disk leading to flickering have repeatedly
been discussed in the literature. For instance, Lyubarskii (\cite{Lyubarskii}),
Yonehara et al.\ (\cite{Yonehara}), Pavlidou et al.\ (\cite{Pavlidou}), and
Dobrotka et al.\ (\cite{Dobrotka10}) developed conceptually similar models 
that differ in details concerning the origin and the propagation of the
disturbances. Of these, the model of Lyubarskii (\cite{Lyubarskii}), 
formulated originally to explain X-ray variations in galactic and 
extragalactic X-ray source, and which invokes viscosity fluctuations 
at different radii, has found most widespread attention. In the
optical range it has sucessfully been applied to MV~Lyr by Scaringi
(\cite{Scaringi14}) [but note that Dobrotka et al.\ (\cite{Dobrotka15})
was able to interpret the same observational data also in the framework of the 
model of Dobrotka et al.\ (\cite{Dobrotka10})].
This scanario is quite attractive, in particular because it seeks
to explain optical flickering in CVs and X-ray fluctuations in neutron star
and black hole binaries as well as AGNs in a unified model. 
However, it has not yet been
convincingly shown that it can explain all flickering characteristics observed 
in CVs (or, alternatively, that certain properties cited in the literature
are based on incorrect assumptions or faulty techniques). Therefore, alternative 
models should not yet be discarded.

None of these mentioned models is concerned with a detailed description
of the emission mechanism that leads to the observed flickering. 
Of the few attempts to construct a realistic physical model for the temporal
and spectral evolution of a flickering flare, the work of Pearson et al.\
(\cite{Pearson}) deserves to be mentioned, which I discuss
in more detail in Sect.~\ref{Fireballs and the case of SS Cyg}. 

As a broad-band phenomenon, in the optical range flickering is apparent from
the blue to the red atmospheric cutoff and, not surprisingly, in different 
photometric bands it is not independent from each other but strongly
correlated (see, e.g.\ Fig.~2 of Bruch 1992). However, depending on the
temporal evolution of the properties of the emitting light source (e.g.\
temperature and opacity), it is conceivable that individual flares do not 
reach their highest intensity simultaneously at all wavelengths, thus 
leading to time lags between the flickering in different
photometric bands. Detection and quantification of such lags will 
provide constraints on physical mechanisms explaining flickering. 

A temporal shift of the flickering at different wavelengths can be traced by
measuring the location of the peak of the cross-correlation function (CCF) of 
light curves in various bands. This technique has been applied by
Jensen et al.\ (\cite{Jensen}), who
observed in this way a lag of about 1\hoch{m} between the flickering
in the optical and X-ray bands in TT~Ari. Similarly, Balman \&
Revnivtsev (\cite{Balman}) detected lags between 96\hoch{s} and 181\hoch{s}
between UV and X-ray observations in five more CVs. 
Using optical observations of AM~Her, Szkody \& Margon (\cite{Szkody}) 
deduced from the asymmetry of the maximum of the CCF a 
lag of the flickering between the $U$ and $B$ bands of 8\hoch{s} in the sense
that variations appear earlier in $U$ than in $B$. On the other hand, 
Bachev et al.\ (\cite{Bachev}) found no conclusive evidence for delays of the 
flickering in different bands of the nova-like variable KR~Aur. This may
be due to the comparatively coarse time resolution of 30\hoch{s} of their
data. Bruch (\cite{Bruch89}, \cite{Bruch92}) also obtained null results, 
but this can probably be attributed to insufficient quality and quantity 
of the observational data.

A more general approach makes use of the coherence function, that is,\ a 
Fourier frequency-dependent measure of the linear correlation between time 
series measured simultaneously in two spectral ranges (Vaughan \& Novak 
\cite{Vaughan}). This permits detecting time lags as a function of the
(temporal) frequency of the underlying signals. This technique has found
widespread application in the study of X-ray binaries and AGNs, but has only
recently successfully been applied to optical data of CVs. Scaringi et
al.\ (\cite{Scaringi13}) studied the Fourier-frequency-dependent 
coherence of light curves of MV~Lyr and LU~Cam observed at high time 
resolution (0\hochpunkt{s}844 -- 2\hochpunkt{s}276) in the SDSS 
bands $u'$, $g'$
and $r'$. They also described in detail the applied data reduction
method. Scaringi et al.\ (\cite{Scaringi13}) found that low-frequency 
variations (below $\approx 10^{-3}$~Hz) occur about 10 seconds later in the
in $r'$ than in $u'$ in LU~Cam. The corresponding time lag is $\approx
3$ seconds in MV~Lyr. At higher frequencies, the errors become too large 
to permit reliable measurement.

Here, I investigate numerous light curves of various CVs observed in 
several photometric systems to measure time shifts of the
flickering as a function of wavelength. As mentioned, the time lags 
observed by Scaringi et al.\ (\cite{Scaringi13}) are most significant for low-frequency variations. Moreover, the (approximate) power-law behaviour of 
the flickering power as a function of frequency (Bruch \cite{Bruch89}, 
\cite{Bruch92})
means that low-frequency variations strongly dominate high-frequency
flickering. Considering furthermore that the CCF will be most sensitive to the
strongest variations as well as the limitations of the available data 
(time resolution, noise, length of individual data sets; see also
Sect.~\ref{Data}), I prefer here to use the cross-correlation function, which is
probably more robust than the much more sophisticated method applied by
Scaringi et al.\ (\cite{Scaringi13}). Not using the coherence function 
implies the loss of some information. But in view of the data limitation,
this disadvantage is compensated by more reliable results than would be 
possible using more sophisticated techniques. To measure 
the time lag of the flickering as a function of wavelength, I
therefore examine 
the CCFs of the light curves, that is,\ the normalized cross
covariant as a function of the shift of the data in a reference band
with respect to the other (comparison) bands of the photometric system. In Sect.~\ref{Reduction method} the reduction method is
explained. The observational data are briefly introduced in 
Sect.~\ref{Data} before the results of this study are presented in
Sect.~\ref{Results}. Scenarios to explain the results are discussed in
Sect.~\ref{Discussion}, and finally, conclusions are drawn in 
Sect.~\ref{Conclusions}.

\section{Reduction method}
\label{Reduction method}

The CCF is calculated in the following way: A light curve is expected to
have been sampled simultaneously in the reference and the comparison 
band in discrete time intervals separated by $\Delta t$. In the data sets
used in this study this is often the case, but sometimes only approximately so.
Then, the light curves are re-sampled by linear interpolation between 
data points at the instances $t_1 + j\, \Delta t$, where $t_1$ is the instance 
of the first data point in the reference band, $j$ is an integer number, and
$\Delta t$ is chosen to be the average time interval between data points in
the original light curve. Comparison and reference data (designated $x$ and
$y$, respectively) are then normalized
to their average value before the mean is subtracted, yielding $n$ pairs of
data points ($x , y$).
The CCF is evaluated at discrete time shifts $\tau_k$, such that 
$\tau_k = k\, \Delta t$ and $k$ is an integral number, 
\begin{displaymath}
CCF(\tau_k) = \frac{\sum_{i=1}^m \left( x_i y_{i+k} \right) \, - \,
              \frac{1}{m}{\sum_{i=1}^m x_i \sum_{i=1}^m y_{i+k}}}
              {\sqrt{\left[ \sum_{i=1}^m x_i^2 -
              \frac{\left(\sum_{i=1}^m x_i \right)^2}{m} \right]
              \left[ \sum_{i=1}^m y_{i+k}^2 -
              \frac{\left( \sum_{i=1}^m y_{i+k} \right)^2}{m} \right]}}
\end{displaymath}
Here, $i = 1 \ldots m$ indicate those data pairs for which reference and 
comparison
data overlap after the comparison data have been shifted in time by $\tau_k$.
The CCF is thus discretely sampled at a time resolution of $\Delta t$.

If reference and comparison data exhibit similar variations but the pattern
of variability is shifted by a time interval ${t_{\rm s}}$ between them, 
the CCF will assume a maximum at $\tau = {t_{\rm s}} \equiv \Delta T_{\rm max}$.

\begin{figure}
   \centering
   \resizebox{\hsize}{!}{\includegraphics{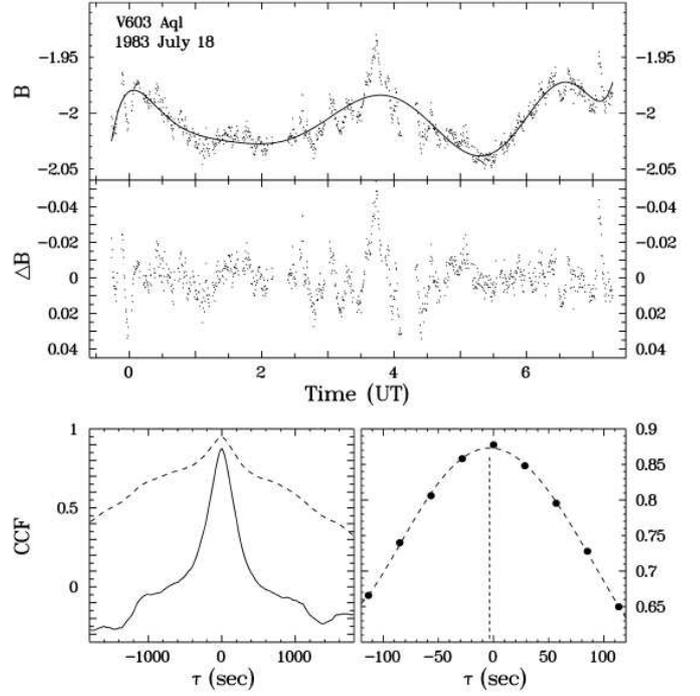}}
   \caption{Example of a light curve and the CCF of the brightness variations
                 in two photometric bands. {\em Top:} $B$-band light curve
                 of V603~Aql observed on 1983 July 18. The solid line is a
                 fit of a high-order polynomial to the data.
                 {\em Centre:} Residuals between the original light curve and
                 the polynomial fit.
                 {\em Bottom left:} CCF of the light
                 curve shown in the top panel with the simultaneously
                 observed light curve in the $U$-band (broken line) and of
                 the residuals between the original data and the polynomial 
                 fit (solid line). {\em Bottom right:}
                 Enlarged section of the maximum  of the CCF shown as a 
                 solid line in the lower left panel
                 (dots) together with a fit of a high-order polynomial 
                 to the data (broken curve). The broken
                 vertical line indicates the location of the maximum of the
                 fitted polynomial.}  
    \label{ccexlres}
    \end{figure}

In principal, it is straightforward to determine the maximum of the CCF
of the data in a reference and a comparison band of light curves
observed simultaneously or quasi-simultaneously at different wavelengths.
However, there are a few details arising from the structure of the individual
data sets available for this study that deserve to be mentioned. As an 
example, the upper panel of Fig.~\ref{ccexlres} shows the $B$ -band light 
curve of the old nova V603~Aql observed in the Walraven photometric system 
on 1983, July 18 (see Walraven \& Walraven, \cite{Walraven}, and Lub \& Pel, 
\cite{Lub}, for the peculiarities of the Walraven system). In
addition to
flickering, it shows another feature frequently observed in CVs, that is,\ 
variations on longer time scales that may not be immediately associated 
with flickering (at least not in the light curve discussed here;
see Bruch \cite{Bruch91}).

The CCF of the light curves in the $B$ band and the simultaneously observed 
$U$ band is shown as a broken line in the lower left frame of 
Fig.~\ref{ccexlres}. The central peak is caused by the correlated
flickering activity in the two bands, but on the whole, the
profile is rather shallow. This is caused by the longer time-scale modulations.
Even if these (in a more general case than for the present example) were a 
manifestation of flickering, the limited length of the light curves -- being
of the same order of magnitude -- would inhibit a reliable measurement of
an associated time lag of the order of a few seconds between different bands.
Therefore, to better isolate the variability due to flickering 
on time scales where a lag can be measured more reliably, it
is appropriate to subtract the slow modulations. This
is done by fitting a polynomial of suitable order to the original data in both
bands, as shown by the solid line in the upper frame of Fig.~\ref{ccexlres}.
In all cases when a polynomial was subtracted, the order was chosen such 
that the polynomial followed the long-term variations satisfactorily, and the
same order was adopted for the reference and the comparison bands.
The light curve in the central fram of Fig.~\ref{ccexlres} 
constitutes the residuals between the
original data and the fitted polynomial in the $B$ band. The solid graph
in the lower left frame represents the CCF of these residual light curves in 
$B$ and $U$. Here, the central peak due to the correlated flickering stands
out much more clearly. 

The lower right frame of Fig.~\ref{ccexlres} contains a magnified version
of the upper part of the CCF maximum (solid dots). 
The coarse resolution reflects
the rather low time resolution of 21\hoch{s} of the example light curve.
To estimate the location of the maximum, a polynomial was fit
to the data of the CCF. Its order depends on the number of available points
in the peak of the CCF (range: $-120\hoch{s} \le \tau \le 120\hoch{s}$ in 
this case). This is the dashed line in the figure. In light curves with a
very high time resolution, the CCF often contains a sharp spike centred on
zero time-shift. It is caused by correlated atmospheric noise in the 
reference and comparison bands and is masked in the fit process. The quality 
of the fit indicates that the maximum of the CCF can be 
determined with high precision by calculating the maximum of the fit 
polynomial. Its location is marked by the broken vertical line. 
The corresponding $\Delta T_{\rm max}$ is slightly negative in this example.

\begin{figure}
   \centering
   \resizebox{\hsize}{!}{\includegraphics{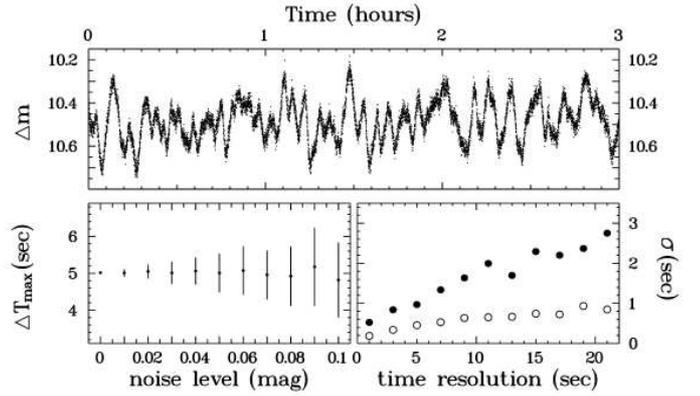}}
   \caption[]{{\em Top:} Artificial flickering light curve used to estimate
             the accuracy of the determination of flickering time lags.
             {\em Bottom left:} Dependence of $\Delta T_{\rm max}$ on the
             noise level. {\em Bottom right:} Expected error of 
             $\Delta T_{\rm max}$ as a function of the time resolution of
             the light curve for low (0\hochpunkt{m}02; open symbols) and
             moderately high (0\hochpunkt{m}05; filled symbols) noise levels.} 
    \label{artlc}
    \end{figure}

To obtain a notion of the accuracy to which $\Delta T_{\rm max}$ can
be measured, a simple shot-noise model was used to generate artificial
flickering light curves. While this may not reproduce the flickering properties
observed at least in some CVs and in many X-ray binaries such as the linear
rms - flux relation, the dependence of the accuray of $\Delta T_{\rm max}$ on
properties such as statistical noise in the data or their time resolution will
not depend strongly on such details. Thus, 
100 artificial light curves were generated, each of
which subsequently served as reference data. One of them
is shown in the upper frame of Fig.~\ref{artlc}. 
The light curves were constructed in the following way: 
1\,500 simulated flares were distributed randomly on a time base of three 
hours, which is the typical duration of the real light curves studied here.
Their amplitudes are distributed between $0\hoch{m} < A \le 0\hochpunkt{m}1$ 
such that the probability of a flare with a particular amplitude to occur 
decays linearly from the minimum to the maximum value of $A$ and the
probability for a flare with the maximum amplitude goes to zero.
The flares are symmetric and rise and decay at a mean rate of 
0\hochpunkt{m}5 per hour, allowing for a random
scatter equally distributed between $0 < {\rm d}m/{\rm d}t \le 1$
magnitudes per hour around this value. The superposition of these 
flares, sampled in 1\hoch{s} intervals, plus an arbitrary constant, 
constitutes the light curves. To make the light curve displayed in
Fig.~\ref{artlc} look more natural, random Gaussian noise with 
$\sigma = 0\hochpunkt{m}01$ was added. Second versions of the same data sets 
were created by applying a constant time shift of 5\hoch{s} to the originals.
These served subsequently as comparison light curves. Using a constant
is acceptable here because a possible
dependence of the time lag on flickering frequency as observed in MV~Lyr
and LU~Cam (Scaringi et al.\ \cite{Scaringi13}) will not have a strong
bearing on the accuracy of the $\Delta T_{\rm max}$ determination, the more
so because the CCF is -- as mentioned -- most sensitive to the stronger
low-frequency variations.

To study the accuracy of $\Delta T_{\rm max}$ on the noise level,
Gaussian noise was added to the reference and the 
comparison light curves before $\Delta T_{\rm max}$ was calculated.
This was repeated for different noise amplitudes.
The results are shown in the lower left frame of Fig.~\ref{artlc}, 
where for each noise level the mean value of $\Delta T_{\rm max}$ derived 
from the 100 trial light curves is plotted. The error bars are the 
standard deviation $\sigma$ around the mean and thus indicate the
accuracy with which $\Delta T_{\rm max}$ can be measured in an individual
light curve. 

The dependence of the accuracy on the time resolution of the data was
determined in a similar way: The artificial light curves were binned to
simulate a lower resolution before noise was added. Then, the standard
deviation of $\Delta T_{\rm max}$ was calculated. It is
shown in the bottom right frame of Fig.~\ref{artlc} as a function of the
time resolution for low (0\hochpunkt{m}02; open circles) and intermediate
(0\hochpunkt{m}05; filled circles) noise levels. 

The accuracy of the results also depends on the length of the light curves
(or, equivalently, on the number of data points). 
Tests with 100 data sets with half (twice) the 
length of the original ones but with otherwise identical characteristics, 
subjected to the same noise levels as shown in the lower left frame of 
Fig.~\ref{artlc}, resulted in errors of $\Delta T_{\rm max}$ that were 
1.44$\pm$0.21 (0.70$\pm$0.06) larger. Thus, not unexpectedly, 
the errors scale to a high degree of accuracy
with the inverse of the square root of the 
length of the light curves.

This exercise shows that $\Delta T_{\rm max}$ can be measured to an
accuracy of the order of a tenth of a second in high-quality (long, low noise, 
high time resolution) light curves, but is limited to several seconds if they 
are shorter, noisier and/or of low time resolution.

\begin{table*}
\caption{Mean flickering time lags (in seconds) in cataclysmic variables}
\label{Mean flickering time lags}

\begin{tabular}{l*{12}{@{\hspace{1ex}}c}}
\hline\hline

\multicolumn{1}{l}{Name} &
\multicolumn{1}{c}{B/W}  &
\multicolumn{1}{c}{B/U}  &
\multicolumn{1}{c}{B/L}  &
\multicolumn{1}{c}{B/V}  &
\multicolumn{1}{c}{B/R}  &
\multicolumn{1}{c}{B/I} \\ 
\hline 

\object{RX And}     & 
 & 
\phantom{0}-3.12\phantom{$\pm$\phantom{0}0.00} (\phantom{0}1) &
 & 
\phantom{0}-5.19\phantom{$\pm$\phantom{0}0.00} (\phantom{0}1) &
 & 
 &
\\
\object{AE Aqr}     & 
\phantom{0}-2.16\phantom{$\pm$\phantom{0}0.00} (\phantom{0}1) &
\phantom{-0}2.59$\pm$\phantom{0}7.97 (55)           &
\phantom{0}-5.93$\pm$17.29           (\phantom{0}2) &
\phantom{0}-2.82$\pm$11.14           (54)           &
\phantom{-0}4.38$\pm$\phantom{0}7.25 (47)           &
\phantom{-0}5.02$\pm$\phantom{0}9.77 (31)           \\
\object{V603 Aql}   & 
\phantom{0}-2.40$\pm$\phantom{0}3.53 (\phantom{0}4) &
\phantom{0}-3.85$\pm$\phantom{0}2.46 (15)           &
\phantom{0}-3.75$\pm$\phantom{0}4.18 (\phantom{0}4) &
\phantom{-0}2.44$\pm$\phantom{0}4.04 (15)           &
\phantom{-}10.58$\pm$\phantom{0}2.92 (\phantom{0}7) &
\phantom{-}17.15$\pm$\phantom{0}7.21 (\phantom{0}2) \\
\object{TT Ari}     & 
\phantom{-0}0.02$\pm$\phantom{0}1.29 (\phantom{0}3) &
\phantom{0}-1.45$\pm$\phantom{0}1.16 (24)           &
\phantom{-0}0.55$\pm$\phantom{0}0.86 (\phantom{0}3) &
\phantom{-0}1.20$\pm$\phantom{0}1.30 (24)           &
\phantom{-0}2.25$\pm$\phantom{0}0.53 (13)           &
 \\
\object{QU Car}     & 
 & 
\phantom{0}-1.65$\pm$\phantom{0}0.82 (\phantom{0}3) &
 & 
\phantom{-0}4.90$\pm$\phantom{0}2.56 (\phantom{0}3) &
 & 
 \\ [1ex]
\object{BV Cen}     & 
\phantom{-0}1.88$\pm$\phantom{0}3.04 (\phantom{0}5) &
\phantom{0}-1.53$\pm$\phantom{0}1.72 (\phantom{0}5) &
\phantom{0}-3.18$\pm$\phantom{0}4.12 (\phantom{0}5) &
\phantom{0}-0.05$\pm$\phantom{0}4.41 (\phantom{0}5) &
 & 
 \\
\object{WW Cet}     & 
\phantom{0}-1.26$\pm$\phantom{0}2.13 (\phantom{0}6) &
\phantom{-0}1.04$\pm$\phantom{0}3.53 (\phantom{0}7) &
\phantom{-0}2.07$\pm$\phantom{0}4.60 (\phantom{0}6) &
\phantom{-0}0.84$\pm$\phantom{0}3.47 (\phantom{0}7) &
\phantom{0}-1.78\phantom{$\pm$\phantom{0}0.00} (\phantom{0}1) &
\phantom{0}-1.77\phantom{$\pm$\phantom{0}0.00} (\phantom{0}1) \\
\object{T CrB}      & 
 & 
\phantom{-0}0.68$\pm$\phantom{0}1.71 (\phantom{0}2) &
 & 
\phantom{-0}0.43$\pm$\phantom{0}3.11 (\phantom{0}2) &
 & 
 \\
\object{SS Cyg}     & 
 & 
\phantom{-0}1.06$\pm$\phantom{0}3.12 (\phantom{0}7) &
 & 
\phantom{-0}4.22$\pm$\phantom{0}1.41 (\phantom{0}6) &
 & 
 \\
\object{HR Del}     & 
 & 
\phantom{0}-0.75$\pm$17.79           (19)           &
 & 
\phantom{-0}4.23$\pm$10.70           (19)           &
\phantom{-0}5.39$\pm$\phantom{0}4.14 (\phantom{0}4) &
 \\ [1ex]
\object{DO Dra}     & 
 & 
\phantom{-0}1.09$\pm$\phantom{0}0.06 (\phantom{0}2) &
 & 
\phantom{-0}2.13$\pm$\phantom{0}0.42 (\phantom{0}2) &
\phantom{-0}1.65$\pm$\phantom{0}0.22 (\phantom{0}2) &
 \\
\object{V795 Her}   & 
 & 
\phantom{-0}1.63$\pm$\phantom{0}0.01 (\phantom{0}2) &
 & 
\phantom{-0}0.45$\pm$\phantom{0}0.58 (\phantom{0}2) &
\phantom{0}-0.07$\pm$\phantom{0}0.00 (\phantom{0}2) &
 \\
\object{RS Oph}     & 
\phantom{-}10.93\phantom{$\pm$\phantom{0}0.00} (\phantom{0}1) &
\phantom{-0}5.10$\pm$\phantom{0}5.28 (\phantom{0}8) &
\phantom{0}-3.28$\pm$\phantom{0}0.19 (\phantom{0}2) &
\phantom{-0}3.96$\pm$\phantom{0}3.29 (\phantom{0}8) &
\phantom{-0}8.29$\pm$\phantom{0}0.97 (\phantom{0}2) &
\phantom{-}10.20$\pm$\phantom{0}1.37 (\phantom{0}2) \\
\object{V426 Oph}   & 
\phantom{-0}2.87$\pm$\phantom{0}1.96 (\phantom{0}3) &
\phantom{-0}0.67$\pm$\phantom{0}2.37 (\phantom{0}3) &
\phantom{-0}1.63$\pm$\phantom{0}0.62 (\phantom{0}3) &
\phantom{-0}0.67$\pm$\phantom{0}1.31 (\phantom{0}3) &
 & 
 \\
\object{CN Ori}     & 
 & 
\phantom{0}-5.64\phantom{$\pm$\phantom{0}0.00} (\phantom{0}1) &
 & 
\phantom{-0}0.94\phantom{$\pm$\phantom{0}0.00} (\phantom{0}1) &
\phantom{-0}4.93\phantom{$\pm$\phantom{0}0.00} (\phantom{0}1) &
 \\ [1ex]
\object{RU Peg}     & 
 & 
\phantom{-0}2.61\phantom{$\pm$\phantom{0}0.00} (\phantom{0}1) &
 & 
\phantom{-0}5.28\phantom{$\pm$\phantom{0}0.00} (\phantom{0}1) &
 & 
 \\
\object{IP Peg}     & 
 & 
\phantom{0}-1.15\phantom{$\pm$\phantom{0}0.00} (\phantom{0}1) &
 & 
\phantom{0}-1.17$\pm$\phantom{0}0.57 (\phantom{0}2) &
\phantom{0}-0.08$\pm$\phantom{0}0.49 (\phantom{0}2) &
 \\
\object{GK Per}     & 
 & 
\phantom{-0}0.19$\pm$\phantom{0}2.66 (15)           &
 & 
\phantom{-0}1.97$\pm$\phantom{0}2.93 (15)           &
\phantom{-0}2.46$\pm$\phantom{0}2.86 (12)           &
 \\
\object{AO Psc}     & 
 & 
\phantom{0}-0.39$\pm$\phantom{0}0.98 (11)           &
 & 
\phantom{-0}0.44$\pm$\phantom{0}1.11 (11)           &
\phantom{-0}0.19$\pm$\phantom{0}0.94 (11)           &
 \\
\object{VY Scl}     & 
\phantom{0}-0.15$\pm$\phantom{0}0.83 (\phantom{0}4) &
\phantom{-0}0.65$\pm$\phantom{0}1.21 (\phantom{0}4) &
\phantom{-0}0.50$\pm$\phantom{0}1.90 (\phantom{0}4) &
\phantom{-0}2.66$\pm$\phantom{0}0.39 (\phantom{0}4) &
 & 
 \\ [1ex]
\object{RW Tri}     & 
 & 
\phantom{-0}6.16$\pm$\phantom{0}3.80 (\phantom{0}4) &
 & 
\phantom{-0}4.35$\pm$\phantom{0}3.84 (\phantom{0}4) &
\phantom{-0}8.81$\pm$\phantom{0}6.55 (\phantom{0}4) &
 \\
\object{UX UMa}     & 
 & 
\phantom{-}15.74\phantom{$\pm$\phantom{0}0.00} (\phantom{0}1) &
 & 
-13.89\phantom{$\pm$\phantom{0}0.00} (\phantom{0}1) &
-30.96\phantom{$\pm$\phantom{0}0.00} (\phantom{0}1) &
 \\
\object{IX Vel}     & 
 & 
\phantom{-0}0.21$\pm$\phantom{0}2.48 (\phantom{0}5) &
 & 
\phantom{0}-5.60$\pm$\phantom{0}8.23 (\phantom{0}5) &
 & 
 \\
\hline
\end{tabular}
\end{table*}
\section{Data}
\label{Data}

A total of 197 light curves of 23 different CVs were used in this study.
They were observed in several photometric systems over an interval of about 
two decades at different telescopes by many observers with a variety of 
scientific purposes in mind. They are thus quite 
heterogeneous. The data can roughly be grouped and characterized as follows:

$VBLUW$ (Walraven) light curves: They were obtained by various observers at
the Dutch telescope at ESO/La Silla, equipped with the Walraven photometer
(Walraven \& Walraven, \cite{Walraven}). 
The instrument permits only a limited choice
of configurations, resulting in a time resolution of about 21\hoch{s} of the
present data, which may be at the limit of usefulness for the current purpose.
The five passbands with decreasing wavelengths from V in the visual to W in the
ultraviolet are observed simultaneously. 

$UBV$ light curves: Data in this photometric system were generally observed
sequentially in the three bands. This also inhibited a high time resolution. 
Most of the $UBV$ light curves regarded here are limited to a resolution of
$\approx$15\hoch{s} or lower.

$UBVRI$ light curves: Many suffer from the same disadvantages as their $UBV$
counterparts, resulting thus in a relatively coarse time resolution. However, 
some have been obtained using a photometer equipped with a rapidly rotating 
filter wheel (Jablonski et al.\ \cite{Jablonski}), 
which permits obtaining quasi-simultaneous measurements in all 
passbands. These data sets were observed at a time resolution of 5\hoch{s}, 
and are thus more suitable for the present purpose.

$UBVR*$ (Stiening) light curves: 
These are data sets were obtained with the Stiening
photometer (Horne \& Stiening \cite{Horne}), which is equipped with four filters
similar but not identical to the standard $UBVR$ filters. All passbands
are observed strictly simultaneously. The disadvantage that this photometric
system is not well calibrated is by far outweighed in the present context by
the high time resolution used to observe the light curves employed here: 
Most have a sampling interval of 0\hochpunkt{s}5 or 1\hoch{s}, while
some even reach 0\hochpunkt{s}2.

Thus, the most suitable data sets are those observed at high time resolution
in the $UBVRI$ system and, above all, the Stiening data.

\section{Results}
\label{Results}

All available light curves were subjected to the reduction method outlined in 
Sect.~\ref{Reduction method}. The $B$ 
band was used as reference band\footnote{One of the light curves of AE~Aqr
was observed only in $U$, $V$ and $R$.
In this case, the shortest wavelength band served as reference.}. 
The light curves of all other bands were cross-correlated with the $B$ 
-band light curve, and the time lags $\Delta T_{\rm max}$ of the maxima
of the respective CCFs were determined. In some cases, only parts
of the light curves were used. This applies in particular to AE~Aqr, which
sometimes exhibits periods of no or very weak flaring activity that were
then excluded. Eclipses in some light curves of eclipsing CVs were
also removed. Strong long-term variations (time scale:
hours), if present, were removed by subtracting a fitted polynomial of
suitable degree (see Fig.~\ref{ccexlres}). 

Table~\ref{Mean flickering time lags} 
contains average values of 
$\Delta T_{\rm max}$ of the maximum of the CCF of each comparison band
with respect to the reference band as indicated in the table header. 
This is interpreted as the time lags
of the flickering in these bands. Since short data sets should contribute
to the average to a lesser degree than long ones, the results for the 
individual light curves were weighted with their total duration. The number
$n$ of light curves contributing to the mean are given in brackets. The
errors are standard deviations derived from all light curves of a given
star. No
selection was made. All light curves, regardless of the photometric system
or the time resolution, enter the average.

\subsection{The ensemble}
\label{The ensemble}

No clear picture emerges for
the individual systems. In most cases,
the standard deviation is considerable, often significantly larger than the
average value of $\Delta T_{\rm max}$. This is not surprising in view of
the expected errors determined from the simulations performed
in Sect.~\ref{Reduction method} and
the limited quality of most light curves. Moreover, more often than not
(disregarding the errors),
there is no monotonical dependency of $\Delta T_{\rm max}$ as a function of 
wavelength difference between reference and comparison bands, as might be 
naively expected. However, 
when the investigated stars are considered as an ensemble, at least a trend
appears.

\begin{table}
\caption{Isophotal wavelengths}
\label{Isophotal wavelengths}
\vspace{1em}

\begin{tabular}{ll|ll}
\hline
\multicolumn{2}{l|}{UBVRI system} & \multicolumn{2}{l}{Walraven system} \\
Band      & $\lambda_{\rm iso}$ (\AA) & Band & $\lambda_{\rm iso}$ (\AA) \\
\hline
U         & 3573.2                 & W      & 3231.9 \\
B         & 4347.6                 & U      & 3618.2 \\
V         & 5458.6                 & L      & 3832.6 \\
R         & 6454.9                 & B      & 4283.1 \\
I         & 7992.0                 & V      & 5417.9 \\
\hline
\end{tabular}
\end{table}

   \begin{figure}
   \centering
   \resizebox{\hsize}{!}{\includegraphics{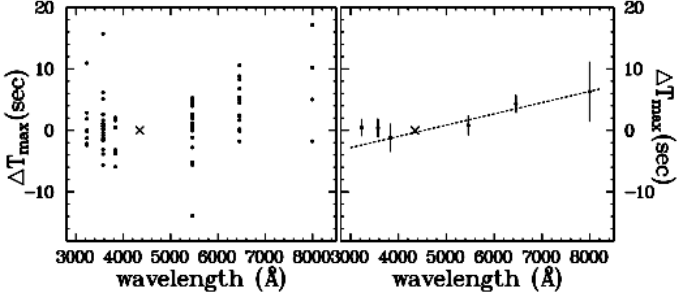}}
   \caption[]{{\em Left:} Average $\Delta T_{\rm max}$ for all investigated 
                CVs as
                a function of isophotal wavelength of the comparison band.
                {\em Right:} Average $\Delta T_{\rm max}$ of all individual light
                curves, weighted in proportion to their duration. The broken
                line is a linear least-squares fit to the data points 
                (excluding the points referring to the $W$ and $U$ bands),
                weighted by the inverse of their standard deviations. The
                error bars represent the mean error of the mean.
                In both frames the location of the reference band ($B$) is
                indicated by a cross.}
    \label{d-deltat}
    \end{figure}

In Fig.~\ref{d-deltat} (left)
the average $\Delta T_{\rm max}$ for each system is shown as a
function of the isophotal wavelength $\lambda_{\rm iso}$ of the comparison band.
$\lambda_{\rm iso}$ was calculated using the transmission curves as listed 
by Bessell (\cite{Bessell}) for the $UBVRI$ system and by Lub \& Pel 
(\cite{Lub}) for the
Walraven system [applying a correction based on the revision of the
mean wavelengths by de Ruiter \& Lub (\cite{deRuiter})], together with a first-order
approximation of the spectral density $S_\lambda$ of a CVs according to the 
law $S_\lambda \propto \lambda^{-7/3}$, valid over a wide wavelength range for
an accretion disk radiating locally as a black body (Lydnen-Bell 
\cite{Lynden-Bell}). The results are listed in 
Table~\ref{Isophotal wavelengths}. Since the transmission curves of the Stiening
system are unknown, $\lambda_{\rm iso}$ was taken to be the same as for the
respective bands of the $UBVRI$ system. Moreover, for simplicity in the case
of the Walraven system the isophotal wavelength of the $U$, $B$ and $V$ bands 
were substituted by those of the $UBVRI$ system in Fig.~\ref{d-deltat}. 
Finally, $\Delta T_{\rm max}$  between
the $B$ and $R$ band of the only light curve of UX~UMa was discarded because
of a very high negative value, deviating strongly from all other results. This
is due to a peculiar and highly asymmetric CCF. The location of the (reference)
$B$ band (where $\Delta T_{\rm max}$ should evidently be zero) 
is indicated by a cross in Fig.~\ref{d-deltat}. 

There is a certain trend for $\Delta T_{\rm max}$  to increase
with wavelength for the photometric bands redward of $U$, meaning that on
average there is a slight offset between the flickering in the blue and in the 
visual or red or infrared in the sense that flickering occurs later at longer
wavelength. But this trend is not continued to the ultraviolet $W$ and $U$
bands. It may not be a coincidence that the points deviating from the linear
relationship refer to those bands that have an isophotal wavelength shorter
than the Balmer limit (see Sect.~\ref{Fireballs and the case of SS Cyg}).

The right frame of Fig.~\ref{d-deltat} is similar to the left one. However,
instead of showing the average $\Delta T_{\rm max}$ values of the respective 
systems, the average of all individual light curves, weighted in proportion to 
their lengths, is shown as a function of the wavelength of the comparison band. 
The error bars represent the mean error of the mean (i.e.\ the
$\sigma/\sqrt{n}$, where $\sigma$ is the standard deviation and $n$ is the
number of data points contributing to the average). The broken line is a 
linear least-squares fit (excluding the $W$ and $U$ bands) 
where the data points were weighted according to the 
inverse of the standard deviation. It has an inclination of 
$(1.9 \pm 0.3) \, 10^{-3}$ sec/\AA\footnote{Weighting alternatively 
the data points 
according to the number of contributing light curves changes this value only 
slightly to $(2.1 \pm 0.4) \, 10^{-3}$ sec/\AA.}. Again, the cross indicates 
the location of the reference band that did not enter the fit.

The trends observed in Fig.~\ref{d-deltat} suggests the consistent 
presence of flickering time lags in CVs, but may not yet be considered 
conclusive evidence. This is the more so because the data points shortward of 
the Balmer jump deviate from the linear relationship. It may be that well-defined time lags in some system with high-quality data are ``diluted'' by
large errors from lower quality light-curves when regarding the entire 
ensemble. This is suggested by more convincing results on some individual 
systems that I discuss subesequently.

\subsection{Individual systems}
\label{Individual systems}

While there is thus a trend for a time lag of the flickering at different
wavelengths when regarding the whole ensemble of investigated light curves,
this is, as mentioned, in general not evident when regarding individual stars. 
There are exceptions, however. Concentrating on high-quality, low-noise light
curves with a time resolution of $\le$5\hoch{s} and using as criteria
(i) that a reasonable number of light curves are available to
render the results statistically reliable, (ii) that the standard deviation of 
the $\Delta T_{\rm max}$ values of the individual curves is significantly smaller
than the difference of the average $\Delta T_{\rm max}$ at short and long 
wavelengths, and (iii) that the average $\Delta T_{\rm max}$ is a monotonic 
function of wavelength and has a different sign for comparison bands shorter 
and longer than the reference band, two systems were identified where 
$\Delta T_{\rm max}$  has a convincing dependence on wavelength. These are 
V603~Aql and TT~Ari. 

\begin{table}
\caption{Restricted results for individual systems}
\label{Restricted results for individual systems}
\vspace{1em}

\begin{tabular}{l@{\hspace{1ex}}lll}
\hline
     &   & \phantom{-}V603 Aql & TT Ari \\ 
\hline
$\Delta T_{\rm max}\, (B/U)$ & (sec) & 
\phantom{0}-4.5$\pm$2.3 (7) &
-2.0$\pm$0.6 (16) \\
$\Delta T_{\rm max}\, (B/V)$ & (sec) & 
\phantom{0-}4.1$\pm$1.8 (7) &
\phantom{-}1.1$\pm$0.8 (16) \\
$\Delta T_{\rm max}\, (B/R)$ & (sec) & 
\phantom{-}10.6$\pm$2.9 (7) &
\phantom{-}2.3$\pm$0.5 (13) \\
$\Delta T_{\rm max}\, (B/I)$ & (sec) & 
\phantom{-}17.2$\pm$7.2 (2) &
\phantom{-}-- \\ [1ex]
d$\Delta T_{\rm max}$/d$\lambda$ & (sec/\AA) &
\phantom{-}(5.0$\pm$0.3)\,$10^{-3}$ &
\phantom{-}(1.5$\pm$0.1)\,$10^{-3}$ \\
\hline
\end{tabular}
\end{table}

\begin{figure}
   \centering
   \resizebox{\hsize}{!}{\includegraphics{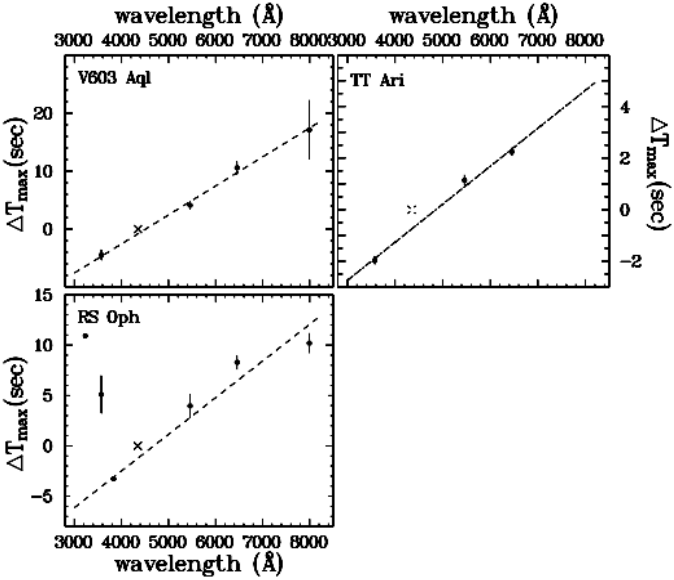}}
      \caption[]{Average $\Delta T_{\rm max}$ derived from high time
                resolution light curves of V603~Aql, TT~Ari and RS~Oph as
                a function of isophotal wavelength of the comparison band.
                The broken
                lines are linear least-squares fits to the data points (in
                the case of RS~Oph without considering the points referring
                to the $W$ and $U$ bands),
                weighted by the inverse of their standard deviations.
                The location of the reference band ($B$) is
                indicated by a cross. The error bars represent the mean error
                of the mean.}
    \label{individu}
    \end{figure}

For both stars, these results are based almost exclusively on the high
time resolution $UBVR*$ light curves. Only for V603~Aql, additionally,
two $UBVRI$ data sets were used. 
Table~\ref{Restricted results for individual systems}
lists the average $\Delta T_{\rm max}$ values 
together with their statistical errors. The number of contributing light 
curves is given in brackets. Figure~\ref{individu} (upper frames), which is 
organized in
the same way as the right frame of Fig.~\ref{d-deltat}, show the same
results graphically. Clearly, $\Delta T_{\rm max}$ strictly obeys a linear
relationship with wavelength $\lambda,$ which, in these cases, continues to
the $U$ band. The derivative of $\Delta T_{\rm max}$ 
with respect to $\lambda$, that is,\ the inclination of a linear fit to
the data points weighted by the inverse of their standard deviation, is
listed in the last line of 
Table~\ref{Restricted results for individual systems}. 

If I relax the criteria formulated in the first paragraph of this
subsection, permitting that in the bands blue-ward of the Balmer jump
$\Delta T_{\rm max}$ deviates from a monotonical relationship with wavelength,
and including also light curves observed with lower time resolution
(but requiring instead that at least four remaining data points are available),
Table~\ref{Mean flickering time lags} 
shows that the relationship
between $\Delta T_{\rm max}$ and $\lambda$ is also quite significant for
RS~Oph (lower left frame of Fig.~\ref{individu}) for wavelengths longer
than the Balmer limit. For this system, the derivative
${\rm d}\Delta T_{\rm max}/{\rm d}\lambda = (3.6 \pm 0.5)\, 10^{-3}$ sec/\AA.  

${\rm d}\Delta T_{\rm max}/{\rm d}\lambda$
differs for the three stars, being more than three times higher in V603~Aql 
than in TT~Ari, with the value for RS~Oph lying in between. This holds true 
for the average, but also for inclinations
derived from the individual light curves. The magnitude of the flickering 
time lags thus depends on the particular star, but within a given system 
it appears to be more or less stable\footnote{In RS~Oph the
scatter of ${\rm d}\Delta T_{\rm max}/{\rm d}\lambda$ of individual light
curves is larger than in V603~Aql and TT~Ari, which may be due to the 
inferior quality of the available data.}. 
However, this statement being based
here on only three stars still requires confirmation from high-quality data of 
other CVs.  

Scaringi et al.\ (\cite{Scaringi13}) found that the time lags in the
stars studied by them depend on the frequency of the fluctuations, which are
larger for variations occuring on longer time scales (and even may change
sign on short time-scales). The CCF technique applied here is not able to
distinguish between different temporal frequencies. Therefore, the observed
time lags should represent some kind of weighted mean over all time scales.
Since the CCF is most sensitive to the strongest variations, the mean is
expected to be biased to the time lags of the dominating fluctuations. Bruch 
(\cite{Bruch89}, \cite{Bruch92}) has shown that the many CVs these occur on 
time scales of the order of 2\hoch{m} -- 4\hoch{m} as measured
from the width of the central peak of the auto-correlation function of their
light curves.

The fact that a time lags between the flickering in different photometric
bands is clearly measured in the two systems with the highest quality data
available for this study [plus at least one additional system (RS~Oph) 
and also the stars observed by Scaringi et al. (\cite{Scaringi13})] 
gives confidence that the trend observed in the
entire ensemble is not accidental, but a strong indication that a similar
time lag is a universal feature of the flickering in CVs.

\section{Discussion}
\label{Discussion}

As mentioned in the introduction, reports about a time lag of the
flickering in different spectral regions have been published earlier, but
the scope and quantity of these studies remains quite limited. In two older
papers and in two more recent studies a positive detection is claimed.

The finding of Szkody \& Margon (\cite{Szkody}) was that in AM~Her the flickering 
in the $V$ band lags that in the $U$ band by 8\hoch{s} is not based on the
measurement of the peak of the CCF, but on its asymmetry. They measured the
lag at the half-intensity point. But it is by no means obvious how
an asymmetry in the peak of a CCF is related to a time lag between the
correlated functions. Indeed, tests with artificial light
curves showed that it is not easy to introduce asymmetries in the CCFs
by shifting the flares in one of the bands or by changing their 
shape\footnote{Even the extreme case where the flares are triangular in
the reference band and saw-tooth-shaped in the comparison band (setting the
rising branch of the original triangle to zero) only introduced a very
slight asymmetry (but changed the location of the maximum significantly).}.
However, such asymmetries can be caused by longer time-scale variations if
these are significantly different in the reference and comparison bands. 
Therefore, it is not clear how reliable the results of Szkody \& Margon
(\cite{Szkody}) are. They interpreted their findings in the framework of the
AM~Her star model, which does not apply to the CVs studied here.

More relevant in the present connection is the study of Jensen et al.\ 
(\cite{Jensen}),
who found a correlations between the flickering in X-rays and the optical
in TT~Ari, the same system where I have detected a highly significant
dependence of $\Delta T_{\rm max}$ on the wavelength in 
Sect.~\ref{Individual systems}. The X-rays lag the optical
variations by about a minute. Jensen et al.\ (\cite{Jensen}) interpreted their
results within a model where X-rays are emitted by a 
corona above the accretion disk. The correlation of variations in the
different bands can then be understood as due to the efficiency of acoustic
and magneto-hydrodynamic transport processes from the disk to the corona. 
However, Jensen et al.\ (\cite{Jensen}) did not investigate this scenario 
quantitatively. Balman \& Revnivtsev (\cite{Balman}) obtained similar results
for additional CVs, but interpreted them
differently: In the framework of a model of propagating 
fluctuations (e.g., Lyubarskii \cite{Lyubarskii}), the lag would be due to 
the travel time of matter from the innermost part of a truncated accretion 
disk to the surface of the white dwarf.

While Szkody \& Margon (\cite{Szkody}) focused on the special
case of a strongly magnetic CV, and Jensen et al.\ (\cite{Jensen}) as
well as Balman \& Revnivtsev (\cite{Balman}) compared
flickering in widely differing wavelength ranges, the study of flickering 
time delays in MV~Lyr and LU~Cam by Scaringi et al.\ (\cite{Scaringi13}) can 
be compared much more directly to the present one. As mentioned, at low 
flickering frequencies (several times $10^{-2}$~Hz as read from their
Fig.~3), they measured a delay of $\approx 10$ sec in
LU~Cam and $\approx 3$ sec in MV~Lyr of the $r'$ band variations with respect 
to those in
the $u'$ band. With the average SDSS filter wavelengths as informed on the
SDSS web page (and neglecting the small difference between average and 
isophotal wavelength), this translates into a time lag of $3.8\, 10^{-3}$ 
sec/\AA\, and $1.1\, 10^{-3}$ sec/\AA, respectively. These numbers are 
quite similar to the time lags observed here, strengthening the case for
such lags to be ubiquitous in CVs.

\subsection{A heuristic model}
\label{A heuristic model}

The time lags observed here in the optical bands are much smaller than
those seen by Jensen et al.\ (\cite{Jensen}) and Balman \& Revnivtsev 
(\cite{Balman}) and have the opposite wavelength
dependence. This is not surprising because the reasons for the lags in the
optical bands probably are quite different from those for the lag 
between the optical and X-ray range. Here, I find that
variations in the blue occur slightly earlier than in the red.
In principle this can be understood if during the development of a flickering
flare the radiation characteristics of the underlying light source change 
such that in the early phases of the flare more short wavelength radiation is 
emitted, and later on, the peak of the emission shifts to the red.

To verify that this leads to results that are compatible 
with the observations, a very simple scenario was investigated.
I stress that this is not meant as a physically realistic model but
just to show that based on simple but sensible assumptions, it is
possible to reproduce the results found in Sect.~\ref{Results}.

For this purpose the light source underlying a flickering flare
was approximated by a black body with a temperature 
$\Theta$\footnote{I chose the symbol $\Theta$ for the temperature to avoid confusion with the symbol $T$ that is used in the quantity 
$\Delta T_{\rm max}$ introduced earlier 
for the time lag of the flickering in different bands.}
that over time $t$ varies according to some function $\Theta(t)$. The emitting
area $\cal A$ of the black body was also considered a function of time: 
${\cal A}(t)$. The radiation
emitted by the black body as a function of wavelength $\lambda$ is then 
given by the Planck function $B_\lambda [\Theta(t)]$ and ${\cal A}(t)$. The 
radiation flux $F(t)$ detected in a passband of a photometric system is thus
\begin{displaymath}
F(t) = {\rm const} \int_{\lambda_1}^{\lambda_2} B_\lambda[\Theta(t)] \, 
       {\cal A}(t) \, \cal{T}(\lambda) \, {\rm d}\lambda,
\end{displaymath} 
where $\cal{T}(\lambda)$ is the transmission function of the photometric 
band, and the integration extends between its upper and lower wavelength 
cutoff. The constant depends on the distance of the radiation source and is
of no relevance in the present context.

In the absence of a specific physical model for the flickering flares,
the functions $\Theta(t)$ and ${\cal A}(t)$ are free parameters. They are chosen
here by trial and error such that the resulting flare profiles and their time 
lag $\Delta T_{\rm max}$ as a function of wavelength are compatible with the
observations. Of course, the choices made here can by no means be
considered as unique. The combination of different functions may well lead 
to similar results. 

   \begin{figure}
   \centering
   \resizebox{\hsize}{!}{\includegraphics{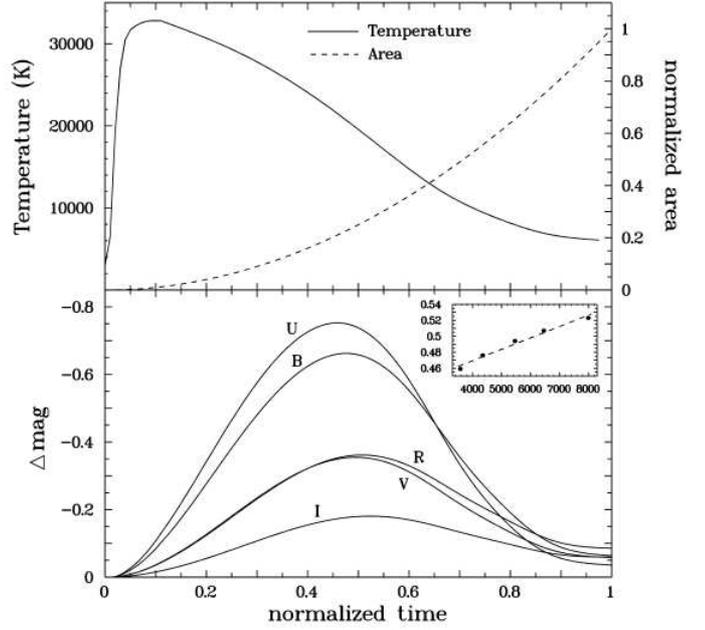}}
      \caption[]{{\em Top:} Assumed temperature (solid line, left-hand scale)
                and emitting area (broken line, right-hand scale) evolution
                of a black body used in simulations of the brightness
                development of a flickering flare as a function of time.
                {\em Bottom:}
                Calculated profiles of a simulated flickering flare in the
                bands of the $UBVRI$ system. The inset shows the times of
                maxima (dots) as a function of the isophotal wavelength of the
                respective bands, together with a linear least-square fit 
                (broken line). See text for details.} 
    \label{flaresim}
    \end{figure}

The time development of the emitting area is taken to be 
${\cal A}(t) \propto t^2$.
This is based on the simple notion of a light source that expands linearly
with time (which may not necessarily be realistic!). With respect to the 
temperature evolution, good results are obtained with a function that
rises rapidly from low temperatures to a maximum of just above 30\,000~K
and then gradually drops to about 6\,000~K at the end of the flare. Both
${\cal A}(t)$ (right-hand scale; normalized to a maximum value of 1; broken
line) and $\Theta(t)$ (left-hand scale; solid line) are shown as a 
function of time (the flare duration being 
normalized to $t_{\rm fl}=1$) in the upper frame of Fig.~\ref{flaresim}.

With this choice of temperature and area evolution, and using the transmission
functions of Bessell (\cite{Bessell}), $F(t)$ (i.e.\ the 
flare light curve) was calculated for the 
passbands of the $UBVRI$ system. The constant was chosen such that $F(t)$
obtains a maximum of $F(t)=1$ in the $U$ band. 
To mimic a constant background
light, a quantity $F_{\rm bg}=1$ was added to all light curves before
transforming them into magnitudes with an arbitrary zero point equal to 
the magnitude of the first point in the light curve corresponding to each
passband. These light curves, depicting the magnitude development of the
flare, are shown in the lower frame of Fig.~\ref{flaresim}.

The simulated flare has several characteristics in common with real flickering
flares: (i) Bruch (\cite{Bruch92}) found the mean ratio of the gradients of 
the rise and decline of flickering flares in the $B$ band of many light 
curves (measured between the base and the peak of the flare) to be 
$1.18 \pm 0.07$, meaning that on average the rise is slightly more rapid 
than the decline. In the simulated flare a very similar gradient ratio of 
1.21 is measured. But note that this value depends quite strongly on the 
choice of $\Theta(t)$ and ${\cal A}(t)$. 
(ii) The amplitude of the flare is highest in the $U$ 
band and declines monotonically in longer wavelength bands\footnote{The
similarity of the amplitudes in $V$ and $R$ is a consequence of the 
significantly larger width of the $R$ passband.}. This agrees with 
observations. However, it is more difficult to make a quantitative assessment 
here because the amplitude depends on the amount of the constant background 
light, which was (unrealistically) assumed to be the same in all passbands in 
the current simulations. (iii) There is a clear dependence of the time of 
maximum on the passband. This is shown in the inset in the lower frame of
Fig.~\ref{flaresim}, where the maximum times (dots) are plotted as a function
of the isophotal wavelength of the photometric band. The broken
line is a linear least-squares fit to the data.
Thus, there is a time lag of the flickering observed at different 
wavelengths. The least-squares fit
yields ${\rm d}\Delta T_{\rm max}/{\rm d}\lambda = 1.4 \times 10^{-5}$ time 
units/\AA. To estimate $t_{\rm fl}$ , the average width 
$\sigma$ of a Gauss fit to the central peak of the auto correlation function
of the $B$ band
of the light curves of TT~Ari used to derive the results quoted in 
Sect.~\ref{Individual systems} was measured: 
$\sigma = 145\hoch{s} \pm 23\hoch{s}$.
Taking this as the typcial flare duration (i.e.\ $t_{\rm fl} \equiv \sigma$), 
${\rm d}\Delta T_{\rm max}/{\rm d}\lambda = 2.1 \times 10^{-3}$
sec/\AA. This is of the same order of magnitude as the observed time lag
for TT~Ari (Table~\ref{Restricted results for individual systems}).

Thus, the simple scenario explored here can explain important characteristics
of the flickering, in particular the main result of the present study, namely
the observed time lag of the flickering at different optical wavelengths.
The numerical value of ${\rm d}\Delta T_{\rm max}/{\rm d}\lambda$ is of the 
same order of magnitude
as that measured in real light curves and depends on the choice of the
temperature and emission area evolution. Although not investigated here, it
may be expected that deviations from black-body radiation have similar
effects. 

The results of this exercise hold for a single flare. But what about an
entire light curve composed of many flares? If flickering were
to consist of 
the additive superposition of many independent flares, it would be easy to
construct such a light curve and investigate its properties. However, there
are strong indications in CVs (e.g.\ Scaringi et al. \cite{Scaringi12b}) and 
even more so in X-ray binaries (e.g.\ Uttley et al. \cite{Uttley05}) that
flickering is the result not of an additive, but of a multiplicative process.
The origins of disturbances in CV system that gives rise to flickering are
largely speculative, and the physical mechanisms of how they lead to the
emission of optical radiation are unknown. Therefore, the properties of an
ensemble of (nonlinearly) superposed flares are not straightforward to predict, 
and I refrain from investigating this issue further.

\subsection{Fireballs and the case of SS Cyg}
\label{Fireballs and the case of SS Cyg}

The heuristic model discussed in the previous subsection can obviously not
substitute a physical model for flickering flares. It was meant to show
that parameter combinations for a light source exist that can plausibly
explain the observed time lag of the flickering as a function of wavelength.
The evolution of the (black body) temperature and the emitting area during
the flare event was adjusted such that the flicker shape resembles
that observed in real light curves.

A more sophisticated model must take into account the physical conditions in
the flare light source. An interesting attempt in this direction has been
published by Pearson et al.\ (\cite{Pearson}). 
They modelled the light source as a fireball, 
that is,\ a hot, spherically symmetric expanding ball of gas, and calculated
the emitting spectrum as a function of the input parameters and of time,
permitting them to construct light curves at various wavelengths. They
compared the results with light curves (time resolution: 2\hoch{s}) in six 
bands between 3590~\AA\, and 7550~\AA\, of a particular flare extracted from 
high-speed spectro-photometric 
observations of the dwarf nova SS~Cyg. Adjusting the model parameters such
that the observed light curves are well fitted yields encouraging results if
the fireball is assumed to be in isothermal expansion.

\begin{figure}
   \centering
   \resizebox{\hsize}{!}{\includegraphics{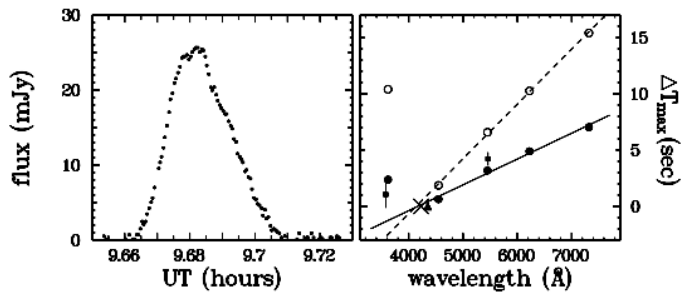}}
      \caption[]{{\em Left frame:} A flickering flare of SS~Cyg centred on
                4545~\AA\, on 1998, July 8, as observed by Pearson et al.\ 
                (\cite{Pearson}).
                {\em Right frame:} $\Delta T_{\rm max}$ as a function of 
                wavelength
                of the comparison band for the observational data (filled 
                circles) and the preferred fireball model (open circles). 
                The solid and broken lines 
                represent a linear
                least-squares fit to the data points (shortest wavelength 
                point excluded). The location of the reference band is
                indicated by a cross. The dots with error bars are
                the average $\Delta T_{\rm max}$ values of SS~Cyg taken from 
                Table~\ref{Mean flickering time lags},
                with the triangle representing the location
                of the respective reference band.}
    \label{lc-sscyg}
    \end{figure}

To investigate how the results of Pearson et al.\ (\cite{Pearson}) 
fit in with the present study, I extracted the light curves of SS~Cyg 
(consisting just of a single flare) from their Fig.~14. For convenience, 
the one corresponding to the
band centred on 4545~\AA\, is shown in the left panel of Fig.~\ref{lc-sscyg}.
In the same way as has been done for the other systems investigated in this
study, the light curves of SS~Cyg were cross-correlated, using the one centred
on 4225~\AA\, as reference. The time lag $\Delta T_{\rm max}$  with respect 
to the reference band is shown as a function of wavelength in the right panel 
of Fig.~\ref{lc-sscyg} (filled circles). The wavelength of the reference band 
is marked by a cross. 

For comparison, the average $\Delta T_{\rm max}$ values for SS~Cyg from 
Table~\ref{Mean flickering time lags}
are shown in the figure as small dots with error bars (mean errors of
the mean). The triangle 
indicates the location of the respective reference band. Unfortunately,
Table~\ref{Mean flickering time lags} 
only contains data of SS~Cyg taken in the $UBV$ system, defining thus but 
two points in the diagram. Therefore, no strong conclusions can be drawn
from the comparison with the
results extracted from Fig.~14 of Pearson et al.\ (\cite{Pearson}). However, 
both data sets appear to be roughly compatible with each other.

Except for the shortest wavelength band centred on 3615~\AA,\, the data points
from Pearson et al. (\cite{Pearson}) 
follow a linear relationship (solid line) with an 
inclination of $(2.3 \pm 0.1) \, 10^{-3}$ sec/\AA. This is very similar to 
the behaviour of V603~Aql, TT~Ari, and RS~Oph 
(Table~\ref{Restricted results for individual systems}
and Fig.~\ref{individu}).

I also determined the times of maximum light of the preferred model of
Pearson et al.\ (\cite{Pearson}) by fitting a high-order polynomial to the 
peak of the model curves extracted from their Fig.~14. The corresponding 
$\Delta T_{\rm max}$ values are shown as open
circles in Fig.~\ref{lc-sscyg}. They follow a linear relationship 
(disregarding the data point at 3615~\AA; broken line) with an inclination 
of $(4.86 \pm 0.01) \, 10^{-3}$ sec/\AA. This is twice the value found
for the observational data, indicating a slight disagreement between model
and observations. 

Although the fireball model is thus quite attractive to explain the time lag
of the flickering as a function of wavelength, a vexing question remains:
Why does the data point at Pearson et al.'s (\cite{Pearson}) 3615~\AA\, band
deviate from the linear relationship between $\Delta T_{\rm max}$ and 
$\lambda$? Or, more generally, how can the tendency for a deviation from
such a relationship at wavelengths shorter than the Balmer limit 
(Fig.~\ref{d-deltat}) be
explained?  This can be due to optical-depth effects in the flaring
light source. Pearson et al.\ (\cite{Pearson}) found that their fireball is
initially dominated by flux coming from optically thick regions. This 
changes in the course of its evolution, so that at late times, optically
thin emission dominates. Thus, more Balmer continuum emission can escape from
the light source at later phases, strengthening the flare at ultraviolet
wavelengths and reversing the $\Delta T_{\rm max}$ -- $\lambda$ relationship
observed at longer wavelengths. 
The balance between optically thick and thin radiation will depend on the
detailed structure of the fireball. Therefore, the late dominance of the
Balmer continuum emission may be seen in some, but not all cases, explaining
the different behaviour of the flickering time lag at very short wavelength
in individual systems (e.g.\ V603~Aql and TT~Ari vs.\ RS~Oph and SS~Cyg). 

\subsection{Alternative scenarios}
\label{Alternative scenarios}

Scaringi et al.\ (\cite{Scaringi13}) sought to explain the time lags along
lines different from fireballs (Sect.~\ref{Fireballs and the case of SS Cyg})
or an evolving black-body-like light source, as
investigated in Sect.~\ref{A heuristic model}. They discarded explanations 
in the context of models of flickering caused by the inward propagation of 
perturbations within the accretions disk [see e.g.\
Lyubarskii (\cite{Lyubarskii}), Pavlidou et al.\ (\cite{Pavlidou}),
Kotov et al.\ (\cite{Kotov}) and Ar\'evalo \& Uttley (\cite{Arevalo}) for 
detailed models of this kind for X-ray binaries and galactic 
nuclei, and Yonehara et al.\ (\cite{Yonehara}), 
Dobrotka et al.\ (\cite{Dobrotka10}) and Scaringi (\cite{Scaringi14}) for CVs], which assume that the observed optical variations are 
directly related to the inward propagation of matter in the accretion disk
because then flickering variations in the blue should lag those in the red. 
This is contrary to what is observed.
They also rejected a scenario of (practically) instantaneous reprocessing of
light of a variable continuum source close to the centre of the accretion 
disk by some structure farther out. The time lag should then basically be
equal to the light travel time. However, the typical dimensions of CVs are
such that only the smallest observed time lags would not exceed the light
travel time from the centre to the periphery of the disk. Moreover, the 
emission at long wavelengths should then be dominated by reprocessed light,
which may not be likely.

If, however, reprocessing does not occur instantaneously, but on the local
thermal time scale Scaringi et al.\ (\cite{Scaringi13}) are able to 
qualitatively reconcile the observed time lags with reprocessing sites not
too far out, provided that only the surface layer of the accretion disk
reprocesses photons from the variable light source. As a final scenario, 
Scaringi et al.\ (\cite{Scaringi13}) mentioned
reverse shocks in the accretions disk (Krauland et al.\ \cite{Krauland})
that may originated in the boundary layer and deposit energy first in the 
hotter inner parts of the disk and then in the cooler outer regions. 

None of scenarios mentioned in the previous paragraph
is backed by a more detailed physical model. Therefore,
it is hard to say if they lead to time scales compatible with the observed
lags and are in accordance with other characteristics of the flickering such 
as, for example, the temporal and spectral development of individual flares. 
But it would be premature to discard them before an attempt has 
been made to predict the implied emission properties in some detail and to 
compare them with observations. It appears that the fireball model 
is at present the only physical model that has been advanced to a stage
where definite predictions can be made
about at least some flickering properties and that realistic parameter 
combinations can be found such that these predictions comply with 
observations. It must be recognized, however, that the model does not
explain what makes the fireball explode in the first place. Moreover, 
it is not known to which degree a (linear or nonlinear) superposition of
individual fireball events can reproduce the statistical properties of 
observed flickering light curves. Thus, it is
by no means clear, which of the mentioned scenarios, if any, is realized
in nature.

\section{Conclusions}
\label{Conclusions}

Confirming earlier results of Scaringi et al.\ (\cite{Scaringi13}),
it has been shown conclusively in the present study that
flickering in cataclysmic variables does not occur exactly simultaneously in
different photometric bands across the optical range. Instead, there is a
time delay in the sense that individual events of the flickering develop
slightly later at red than at blue wavelengths. In many of the
investigated CVs, this effect can only be seen statistically in the ensemble 
of the available data because most of the light curves used are not of 
sufficient quality to resolve the time delay well enough
in particular data sets. However, in the systems with the best data, 
V603~Aql and TT~Ari, the effect can convincingly be measured individually.
This is also true for RS~Oph and  SS~Cyg. To these systems MV~Lyr and 
LU~Cam, observed by Scaringi et al.\ (\cite{Scaringi13}), can be added.

The time lag of the flickering as a function of wavelengths is of the
order of a couple of milliseconds per \AA ngstrom. The exact value may
depend on the particular system, but appears not to change much within the
same star on time scales of several years (i.e.\ the period spanned
by the observational data; this does, of course, not preclude time lag
changes when systems like TT~Ari go into a low state). 

The observed effect can be explained if the evolution of the
radiation characteristics  of the light source(s) responsible for the 
flickering is such that individual flares reach their peak emission
slightly earlier at blue than at red wavelengths. It is not difficult to
construct simple scenarios that are able to reproduce the observed time
lag not only qualitatively, but also quantitatively, and which also agree
with other properties that are generally observed in the flickering.
While such scenarios lead to the observed effects, they can by no means 
substitute a realistic physical mechanism. The fireball model of Pearson 
et al. (\cite{Pearson}) is most advanced when it comes to the actual
radiation emission process that may be seen as flickering. It can reasonably 
well reproduce a flickering flare observed in SS~Cyg in various wavelengths
ranges not only with respect to the wavelength dependent time lag, but also
with respect to the spectral and temporal development. However, 
the detailed emission processes implied in the context of other scenarios 
deserve to be investigated as well. This hold true especially for models 
where, contrary to the fireball model, a better understanding of the origin 
of underlying disturbance exists, for instance,\ the fluctuating disk model. 
It may
be interesting to investigate whether the basic fireball radiation physics 
can be applied to these models as well.

\begin{acknowledgements}
I am deeply indebted to all those colleagues who put their light cuves of
cataclysmic variables at my disposal. For the current study I used data
contributed by N. Beskrovanaya, A.\ Hollander, R.E.\ Nather, M.\ Niehues,
T.\ Schimpke, and N.M.\ Shakovskoy.
I am particularly grateful to R.E.\ Robinson and E.-H.\ Zhang for their 
excellent light curves observed in the Stiening system, which were decisive 
for the success of this work. I thank the referee,\ Simone Scaringi,
for many critical comments that helped to improve this publication.
\end{acknowledgements}


\listofobjects
\end{document}